\documentclass[11pt]{article}

\usepackage{caption}
\usepackage{epsfig}
\usepackage{color}
\usepackage{psfrag}
\usepackage{verbatim}
\usepackage{amsopn}
\usepackage{appendix}
\usepackage{dsfont,amsmath,amssymb,color,units,pstricks}
\usepackage{amsfonts}
\usepackage{graphicx}
\usepackage{hyperref}
\usepackage{array}
\usepackage{lscape}
\usepackage{fullpage}
\usepackage{tikz}
\usepackage{url}
\usepackage{subcaption}

\allowdisplaybreaks
\usetikzlibrary{matrix}

\newcommand{\Ex}{{\mathbb E}}
\newcommand{\PP}{{\mathbb P}}

\newcommand\be{\begin{equation}}
\newcommand\bea{\begin{eqnarray} \nonumber }
\newcommand\ee{\end{equation}}
\newcommand\eea{\end{eqnarray}}

\begin{document}

\unitlength = 1mm
\title{Order statistics of horse racing and the randomly broken stick}

\author{Peter A. Bebbington$^{1,2}$ and Julius Bonart$^{3,4}$}

\maketitle

\noindent\small{$1$: Department of Physics and Astronomy, University College London, London WC1E 6BT; $2$: Trium Capital LLP, 60 Gresham St, London EC2V 7BB; $3$: Financial Computing \& Analytics, Department of Computer Science, University College London, London WC1E 6BT; $4$: CFM--Imperial Institute of Quantitative Finance, Department of Mathematics, Imperial College, London SW7 2AZ}\\ 

\begin{abstract}
  We find a remarkable agreement between the statistics of a randomly divided interval and the observed statistical patterns and distributions found in horse racing betting markets. We compare the distribution of implied winning odds, the average true winning probabilities, the implied odds conditional on a win, and the average implied odds of the winning horse with the corresponding quantities from the ``randomly broken stick problem''. We observe that the market is at least to some degree informationally efficient. From the mapping between exponential random variables and the statistics of the random division we conclude that horses' true winning abilities are exponentially distributed.
\end{abstract}

\section{Introduction}\label{sec:intro}

From time to time nature has a taste for simplicity. It can then be promising to treat unknown variables as purely random, using statistics that are compatible with the constraints, symmetries, or boundary conditions of the given problem, but otherwise as simple as possible. Heavy nuclei are an example of such a system; they are seemingly hopelessly complex, yet the spacings between their energy levels follow well-known statistics of random matrix eigenvalues \cite{Wigner1955, Brody1981}. More recently, one of such statistics, the Marchenko-Pastur distribution, has been found in fluctuations of financial covariance matrices \cite{Laloux1999}, despite the strong non-Gaussian dependencies observed in real financial time series \cite{Bouchaud2009}. Latter example underscores the success of econophysics: Socio-economic human systems are highly non-linear \cite{Toth2011, Donier2015, DonierBonart15b, diMatteo2005} and chaotic \cite{Patzelt2013}, but methods borrowed from statistical physics can still be successful in describing bulk statistics of these systems.

Traditionally, econophysics has somewhat neglected a certain type of financial markets: Betting markets. This is perhaps surprising because economists, on the contrary, have studied betting markets extensively, considering them as a controlled experiment for market efficiency \cite{Williams1999,Figlewski1979,Divos2014}, a key concept in financial economics. Because the outcome of the bet -- win or lose -- is definitely known after a certain time, it is straightforward to draw conclusions from the discrepancy between the implied market odds\footnote{We use here ``implied odds'' in the sense of ``implied probability''.} and the true winning probability. If the difference is large, the market is regarded as inefficient, because its participants are not able to ``price'' the bet correctly. If the difference is small the market is regarded as efficient.
How are the implied market odds calculated? Consider for example a horse race with $n = 3$ horses. Assume that after betting $\$$X on the first horse you would get $\$3$X if this horse wins. The ``price'' of a bet on the first horse is thus 3. If the price of a bet on the second horse is 2 and the price of a bet on the third horse is 6, the second horse is the favourite because the market would pay the smallest ratio of the gain (including the original stake) to the stake itself if this horse wins: Its implied winning odds are $1/2$ so that the gambler's average payoff is zero.

The sum of all the implied odds must be roughly one. We know that this is not always exactly true, for example because bookmakers charge the gamblers a small fee, but for the purpose of this study these tiny deviations are not important. Each horse's implied odds thus represent a segment of the unit interval. We do not know much about horses, but we can guess the simplest statistics for these segments: Draw $n-1$ numbers from the uniform distribution and cut the unit interval at each of these numbers. We thus break a stick of unit length randomly into $n$ pieces, each of which represents the winning odds of an individual horse, participating in a race with $n$ horses. The favourite's odds correspond to the largest segment, the second favourite's odds to the second largest segment, and so on.

This letter reports striking similarities between the empirical distribution of implied odds observed in horse racing betting markets and the order statistics of the random division of the unit interval. Moreover, we find that conditional expectations of the true winning probabilities\footnote{Of course we cannot observe an empirical distribution of true winning probabilities but only aggregate statistics, such as, for example, the average winning probability of the favourite horse.} closely follow their corresponding values from the ``broken stick problem'', as well. We therefore conclude that the true winning probabilities of horses behave like random divisions of the unit interval, and that the market follows these statistics in the implied odds. Finally, from the well known mapping between exponential random variables and the statistics of the random division \cite{Holst1980} we make the somewhat vague statement that horses' true ``abilities'' are exponentially distributed: The probability that a horse with ability $X_i$ wins against $n-1$ other horses is then
\begin{equation}\label{eq:exp}
  P_i = \frac{X_i}{\sum_{j=1}^n X_j}\;,
\end{equation}
in that $P_i$ follows the statistics of the broken stick problem if and only if the $X_i$ are exponentially distributed.

Consider the interval $[0,1]$ and divide it randomly into $n$ sub-intervals. The length of the $k$-th largest sub-interval, which we denote here by $z_{(k)}$, has the distribution \cite{Holst1980}
\begin{equation}\label{eq:dis_zk}
  \PP[z_{(k)}>x|n] = \sum_{j=1}^{k-1}{n\choose{j}}\sum_{\ell=0}^{n-j}(-1)^{\ell-1}{{n-j}\choose{\ell}} [1 - (j+\ell)x]_+^{n-1} + \sum_{\ell=1}^n(-1)^{\ell-1} {n\choose{\ell}}[1 - \ell x]_+^{n-1}\;,
\end{equation}
with $a_+ = \max[a,0]$. We want to compare $\PP[z_{(k)}>x]$ to the empirical distribution of implied winning odds of horse racing betting markets.

We use data collected through Betfair on 12736 races occurring across the British Isles in the period from 31/12/2011 to 15/12/2012. The average number of horses per race is 8.95. We consider only races with at least $5$ horses which reduces the total number of races in our dataset to 11925. Gamblers exchange bets on horses in a limit order book. Sell orders match buy limit orders specified by volumes and \emph{lay decimal odds}. Buy orders match sell limit orders specified by volumes and \emph{back decimal odds}. Decimal odds quote the ratio of the payout amount, including the original stake, to the stake itself. The highest back quote is \emph{larger} than the lowest lay quote\footnote{Note that here buy orders match at \emph{lower} quotes than sell orders.}. The implied winning odds are defined as the reciprocal of the last matched quote before the race starts.

\begin{figure}
  \centering
  \includegraphics[width=0.6\textwidth]{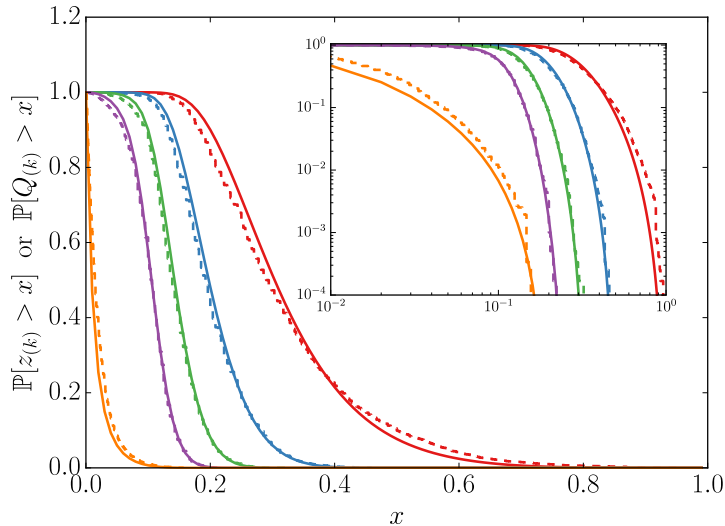}
  \caption{(Dashed) ECCDF of the four favourite and longshot horses' implied odds (red: favourite horse, blue: second favourite, green: third favourite, violet: forth favourite, orange: longshot) and (solid) the cumulative distribution of the corresponding segments of the division of the unit interval, displayed in (main) linear and (inset) double logarithmic scale. Note that there are \emph{no} free fitting parameters.}
  \label{fig:comparison}
\end{figure}

Consider now the implied odds of the $k$-th favourite horse, which we denote by $Q_{(k)}$. Fig. \ref{fig:comparison} compares the ECCDF $\PP[Q_{(k)}>x]$ with the theoretical prediction $\PP[z_{(k)}>x]$ for the favourite, 2nd favourite, 3rd favourite, 4th favourite, and the horse with the least implied winning odds (the ``longshot''), averaged over the number of horses in each race. The agreement is striking and calls for further investigation.  

To compare the true winning probabilities to the order statistics of the broken stick problem we need to calculate average quantities. Table \ref{tab:zQP} shows the expected length of the $k$-th largest segment, the average empirical implied odds of the $k$-th favourite, and the average observed true winning probabilities of the $k$-th favourite, denoted by $P_{(k)}$, for all races in our dataset and for three subgroups containing roughly an equal number of horses: races with $5\le n\le 7$ horses, races with $8\le n\le10$ horses, and races with $n\ge 11$ horses. The theoretical expectation of the segment lengths are calculated by taking the first moment of Eq. \ref{eq:dis_zk} (analytically given in Eq. \ref{eq:zk} below) and averaging over the empirical distribution of $n$.
\begin{table}
  \centering
  \begin{tabular}{p{3.2cm}|c|c|c|c|c}
    \hline
    & favourite & 2nd favourite & 3rd favourite & 4th favourite & longshot\\
    & $k=1$ & $k=2$ & $k=3$ & $k=4$ & $k=n$ \\
    \hline
    $\Ex[Q_{(k)}|n\ge 5]$ & 0.3208 & 0.2001 & 0.1420 & 0.1037 & 0.0210 \\
    $\Ex[P_{(k)}|n\ge 5]$      & 0.3358 & 0.1976 & 0.1345 & 0.0998 & 0.0253 \\
    $\Ex[z_{(k)}|n \ge 5]$ &0.3237 & 0.2046 & 0.1451 & 0.1054 & 0.0157 \\
    \hline
  \end{tabular}
  \begin{tabular}{p{3.2cm}|c|c|c|c|c}
    \hline
    & favourite & 2nd favourite & 3rd favourite & 4th favourite & longshot\\
    & $k=1$ & $k=2$ & $k=3$ & $k=4$ & $k=n$ \\
    \hline
    $\Ex[Q_{(k)}|5\le n\le7]$ & 0.3996 & 0.2399 & 0.1578 & 0.1024 & 0.0336\\
    $\Ex[P_{(k)}|5\le n\le7]$      & 0.4165 & 0.2276 & 0.1503 & 0.0981 & 0.0339\\
    $\Ex[z_{(k)}|5\le n\le7]$ & 0.4081 & 0.2407 & 0.1570 & 0.1012 & 0.0285\\
    \hline
  \end{tabular}
  \begin{tabular}{p{3.2cm}|c|c|c|c|c}
    \hline
    & favourite & 2nd favourite & 3rd favourite & 4th favourite & longshot\\
    & $k=1$ & $k=2$ & $k=3$ & $k=4$ & $k=n$ \\
    \hline
    $\Ex[Q_{(k)}|8\le n\le10]$ & 0.3184 & 0.1985 & 0.1438 & 0.1078 & 0.0182 \\
    $\Ex[P_{(k)}|8\le n\le10]$      & 0.3327 & 0.2081 & 0.1362 & 0.1031 & 0.0233 \\
    $\Ex[z_{(k)}|8\le n\le10]$ & 0.3166 & 0.2041 & 0.1478 & 0.1103 & 0.0128 \\
    \hline
  \end{tabular}
  \begin{tabular}{p{3.2cm}|c|c|c|c|c}
    \hline
    & favourite & 2nd favourite & 3rd favourite & 4th favourite & longshot\\
    & $k=1$ & $k=2$ & $k=3$ & $k=4$ & $k=n$ \\
    \hline
    $\Ex[Q_{(k)}|n\ge 11]$ & 0.2470 & 0.1631 & 0.1247 & 0.1004 & 0.0119 \\
    $\Ex[P_{(k)}|n\ge 11]$      & 0.2614 & 0.1564 & 0.1172 & 0.0977 & 0.0193 \\
    $\Ex[z_{(k)}|n\ge 11]$ & 0.2500 & 0.1703 & 0.1305 & 0.1039 & 0.0065 \\
    \hline
  \end{tabular}
  \caption{Average implied odds and winning probabilities, and expected segment lengths for (from above) all races in our dataset, all races with $n \le 7$, with $8\le n\le10$, and with $n\ge 11$. The theoretical expectation of the segment lengths are calculated by averaging the first moment of Eq. \ref{eq:dis_zk} over the empirical distribution of $n$.}
  \label{tab:zQP}
\end{table}
Not only do the empirical implied odds correspond to the expected segment lengths but the average observed winning probabilities also follow the order statistics of the random division accurately for all horses, with the exception of the longshot. Note that our theoretical estimations of the winning odds based on the expected segment lengths are \emph{parameter free}.

We observe significant discrepancies for the longshot, but the differences between its implied odds, winning probability and segment length are small for races with $5\le n\le 7$ horses and larger for races with more horses. This suggests that gamblers are not able to rank the horses precisely enough when the number of horses is large. Remember that we define the rank of the horse according to the observed implied odds. Therefore, the smallest segment may describe a horse that the market has not recognised as the weakest one. In this case the market's longshot is in reality a slightly stronger horse. This is consistent with the fact that both implied odds and winning probabilities of the longshot are larger than suggested by Eq. \ref{eq:dis_zk}.

We also calculate the implied odds of the $k$-th favourite given that this horse wins. This quantity is naturally larger than the unconditional implied odds of the $k$-th favourite. To find the corresponding theoretical prediction consider the indicator function $I_{(k)}=1$ which is one if a random point in the interval $[0,1]$ lies in the $k$-th largest segment and zero, else.
Then:
\begin{align*}
  \PP[z_{(k)}=x|I_{(k)}=1] = \frac{\PP[I_{(k)}=1|z_{(k)}=x]\PP[z_{(k)}=x]}{\PP[I_{(k)}=1]} = \frac{x \PP[z_{(k)}=x]}{\bar z_{(k)}}\;,
\end{align*}
and
\begin{equation}\label{eq:zkwin}
  \Ex[z_{(k)}|I_{(k)}=1] = \frac{\overline{z^2}_{(k)}}{\bar z_{(k)}}\;.
\end{equation}
Eq. \ref{eq:zkwin} is the theoretical prediction of the $k$-th favourite's odds given that this horse wins.
By using well-known binomial identities we find from Eq. \ref{eq:dis_zk} after a somewhat lengthy calculation\footnote{The identity for $\bar z_{(k)}$ is reported in \cite{David2003}, p. 153, but the authors have no knowledge of a previous appearance of Eq.~\ref{eq:zk2}.} that
\begin{equation}\label{eq:zk}
  \bar z_{(k)} = \frac{1}{n}\sum_{j=k}^n\frac{1}{j} = \frac{1}{n}H_{n,k}\;,
\end{equation}
with the partial harmonic number $H_{n,k}\equiv \sum_{j=k}^nj^{-1}$ and
\begin{align}\label{eq:zk2}
  \overline{z^2}_{(k)} = \frac{2}{n(n+1)}\sum_{j=k}^n\frac{H_{n,j}}{j} = \frac{2}{n+1}\sum_{j=k}^n\frac{\bar z_{(j)}}{j} \;.
\end{align}
Table \ref{tab:win} compares the implied odds of the $k$-th favourite given that it wins with the average length of the $k$-th largest segment given that it contains a random point, see Eq. \ref{eq:zkwin}.
\begin{table}
  \centering
  \begin{tabular}{p{3cm}|c|c|c|c|c}
    \hline
    & favourite & 2nd favourite & 3rd favourite & 4th favourite & longshot\\
    & $k=1$ & $k=2$ & $k=3$ & $k=4$ & $k=n$ \\
    \hline
    $\Ex[Q_{(k)}|{\rm win}]$ & 0.3735 & 0.2148 & 0.1542 & 0.1139 & 0.0886 \\
    $\Ex[z_{(k)}|{I_{(k)} = 1}]$ & 0.3622 & 0.2196 & 0.1549 & 0.1145 & 0.0383 \\
    \hline
  \end{tabular}
  \caption{Average implied odds given that the horse wins and expected segment lengths given that it contains a random point for all races in our dataset (with $n\ge 5$). The theoretical expectation of the segment lengths are calculated by averaging Eq. \ref{eq:zkwin} over the empirical distribution of $n$.}
  \label{tab:win}
\end{table}
We observe again a good agreement between the empirical odds and the theoretical prediction (except for the longshot, see discussion above).

Finally, we calculate the average initial odds of the winning horse which is $0.2148$. Its theoretical prediction follows from Eq. \ref{eq:zkwin},
\begin{equation}
  {\parbox{4.8cm}{\text{Average length of the segment} \\ \text{containing the random point}}} = \sum_{k=1}^n \Ex[z_{(k)}|I_{(k)}=1]\PP[I_{(k)}=1] = \sum_{k=1}^n \overline{z^2}_{(k)} = \frac{2}{n+1}\;,
\end{equation}
which -- after averaging over $n$ -- yields $0.2107$, again very close to the empirical value.

To summarise, we have found a remarkable agreement between the order statistics of the randomly broken stick and the statistical properties of horse racing betting markets. We also  observe that the empirical values of the implied odds and true winning probabilities are close and therefore conclude that this betting market is informationally efficient at least to some degree. Some discrepancies are found for the longshot, because gamblers probably fail to rank the horses accurately when their number is big. Assuming that the implied odds reflect to a large extent the true winning probabilities, we conclude that the ``ability'' of a horse can be defined in such a way that its winning probability is the ratio of its ``ability'' to the sum of all its competitors' abilities, provided ``ability'' is exponentially distributed.

Acknowledgements: Julius Bonart thanks Jean-Philippe Bouchaud, Jonathan Donier and Tomaso Aste for interesting discussions. We would also like to give warm thanks to Peter A. Bebbington's PhD supervisors I. J. Ford and F. M. C. Witte, the funding body EPSRC and the Centre for Doctoral Training in Financial Computing \& Analytics.

\bibliographystyle{unsrt}
\bibliography{bibli.bib}

\end{document}